\documentclass[11pt,twoside]{article}
\usepackage{asp2010}

\resetcounters

\begin{document}

\title{Global-scale Simulations of Stellar Convection and their Observational Predictions}
\author{Benjamin P.\ Brown$^{1,2}$
\affil{$^1$Department of Astronomy, \\
           University of Wisconsin, 475 Charter Street, Madison, WI 53706-1582\\
       $^2$Center for Magnetic Self Organization (CMSO) \\ 
           in Laboratory and Astrophysical Plasmas, \\
           University of Wisconsin, 1150 University Avenue, Madison, WI 53706\\}}

\begin{abstract}
Stars on the lower main sequence (F-type through M-type) have
substantial convective envelopes beneath their stellar photospheres.
Convection in these regions can couple with rotation to build
global-scale structures that may be observable by interferometers that
can resolve stellar disks.  
Here I discuss predictions emerging from 3D MHD simulations
for solar-type stars with the anelastic spherical harmonic (ASH) code
and how these predictions may be observationally tested.
The zonal flow of differential rotation is likely the most
easily observable signature of dynamics occuring deep within the
stellar interior.  
Generally, we find that rapidly rotating suns have a strong solar-like
differential rotation with a prograde equator and retrograde poles
while slowly spinning suns may have anti-solar rotation profiles with
fast poles and slow equators. The thermal wind balance accompanying
the differential rotation may lead to hot and bright poles in the
rapid rotators and cooler, darker poles in slow rotators.  The
convection and differential rotation build global-scale magnetic
structures in the bulk of the convection zone, and these wreaths of
magnetism may be observable near the stellar surfaces.
\end{abstract}

\section{Introduction}
When stars like our Sun are young, they rotate rapidly and have strong
magnetic fields at their surfaces.  Magnetic activity is a nearly
ubiquitous feature of F- to M-type stars on the lower main sequence,
all of which have convective envelopes just below their photospheres.
Younger and more rapidly rotating stars are generally more active
and follow the ``rotation-activity'' relationship
\citep[see e.g.,][]{Pizzolato_et_al_2003}.  Stellar magnetic fields
are thought to arise from dynamo processes occuring in these stellar
convection zones, where turbulent plasma motions couple with rotation
to build global-scale magnetic fields.  At present, stellar dynamo
theory does not explain the observed correlation between rotation and
magnetic activity.  Likewise, our own Sun's eleven year magnetic
activity and sunspot cycles remains a puzzling mystery: despite
intense study, solar dynamo models are at present unable to reliably
predict even large-scale features of the solar cycle.

The nature of solar and stellar dynamos, and the origin of solar
and stellar magnetic fields, remains one of the most important unsolved
problems in stellar astrophysics.  In modern times, the tremendous
growth of computational resources, coupled with detailed pictures of
flows and structure within the solar interior from helioseismology,
has lead to an explosion of dynamo modeling efforts, ranging from
sophisticated 2D mean-field models to fully 3D simulations that can
capture the non-linear dynamics of solar convection
self-consistently.  Indeed it is now possible to model global-scale
convection and dynamo action in the Sun with some fidelity
\citep[e.g.,][]{Brun&Toomre_2002, Brun_et_al_2004,
  Browning_et_al_2006, Miesch_et_al_2006, 
  Miesch_et_al_2008, Miesch&Toomre_2009, 
  Ghizaru_et_al_2010, Racine_et_al_2011}  and those efforts are
being extended to other solar-type stars.  The progress in solar
simulations was helped tremendously by detailed observations of the
Sun which constrained and challenged the simulations.  Detailed
observations of other stars will likewise be necessary for further
progress in a general understanding of stellar convection and dynamo
action, which will itself aid our understanding of the solar interior.

Here I will discuss recent global-scale 3D magnetohydrodynamic 
(MHD) simulations of stellar convection with the anelastic spherical harmonic (ASH) code
\citep{Clune_et_al_1999, Brun_et_al_2004}.  This code has been used to
simulation solar convection and reproduces the observed solar
differential rotation profile relatively well
\citep[e.g.,][]{Miesch_et_al_2006, Miesch_et_al_2008}.  Building on
this success, we have conducted a series of simulations for solar-type
stars rotating more rapidly than the Sun \citep{Brown_et_al_2008} and
have explored the dynamo-generated magnetic fields in several of these
cases \citep{Brown_et_al_2010a, Brown_et_al_2011, Brown_2011}.  These
simulations are beginning to make specific observational predictions
which optical interferometry may be able to test, and here we will
begin laying out what those questions are and how they may be answered.

\section{Simulating stellar convection}
One path towards understanding stellar convection is to conduct
simulations of the plasma motions occuring within the stellar
interior.  Stellar convection spans a vast range of spatial and
temporal scales, which lie well beyond the grasp of direct numerical
simulation even on the largest modern supercomputers.  Consequently, 
models of stellar convection and dynamo action must make various
tradeoffs, either building up from the smallest diffusive scales or
building down from the global-scales.  The later global-scale
simulations will be our focus here, and it is these simulations that can
self-consistently capture the interactions between convection and
rotation in a stratified atmosphere to drive global-scale flows of
differential rotation.  These simulations remain distant from stellar
parameters, and a sense of this gap is given in \cite{Brown_2011}.

\section{Global-scale signatures of convection}
\begin{figure}[!t]
  \begin{center}
    \includegraphics{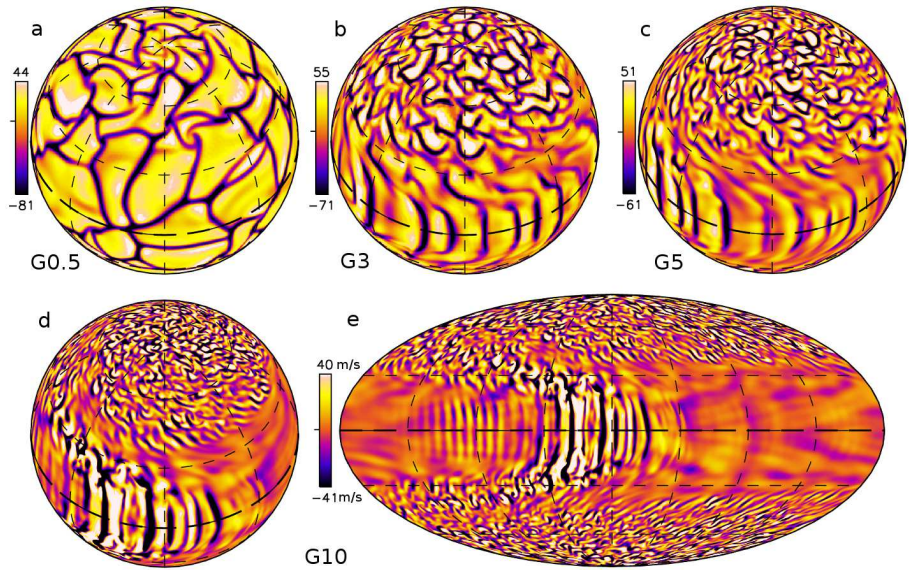}
  \end{center}
  \caption{Patterns of convection in solar-type stars.  Shown are
    radial velocity patterns near the stellar surface for stars
    rotating at $(a)$ $0.5\:\Omega_\odot$, $(b)$ $3\:\Omega_\odot$,
    $(c)$ $5\:\Omega_\odot$, and $(d)$ $10\:\Omega_\odot$.  The broad,
    slow upflows are shown in light tones while the narrow, fast
    downflows are shown in dark tones.  The north pole is visible and
    the equator is denoted by a dashed line.  Clear differences are
    apparent in the polar and equatorial regions, and these become
    more pronounced as the rotation rate increases.  At the highest
    rotation rates, convection near the equator can become confined to
    narrow bands in longitude.  To emphasize this in case~G10 we
    show the whole sphere from $(d)$ in a Mollweide view in $(e)$,
    with equator at middle and poles at top and bottom. These active
    nests of convection retain their identity for many thousands of
    days and propagate at speeds distinct from the stellar rotation
    rate \citep{Brown_et_al_2008}.
    \label{fig:convection}}
\end{figure}

The patterns of convection arising in simulations of solar-type stars
at a variety of rotation rates are shown in Figure~\ref{fig:convection}.  
These simulations span from 0.5 to 10 $\Omega_\odot$, with
$\Omega_\odot$ the current solar rotation rate.  These are labeled
G0.5 through G10 respectively.  When the rotation rate is slow
relative to convective motions (equivalently, the Rossby number is
large) then rotational constraints are weak and convective patterns
are very similar in both polar and equatorial regions.  This is
evident in case~G0.5 (Fig.~\ref{fig:convection}$a$).  As the rotation
rate increases (Rossby number decreases),  significant differences
between the equator and pole emerge.  Convection near the equator
aligns with the rotation axis, forming ``banana-cells'' of
convection.  Near the poles the convection is more isotropic, with
narrow downflow lanes surrounding broader upflows (e.g., case~G5 in
Fig.~\ref{fig:convection}$c$).  As the rotation rate is increased the
horizontal scale of individual convective cells becomes smaller.
Individual global-scale convective cells will likely be nearly
impossible to detect on main-sequence solar-type stars; indeed their
detection has eluded helioseismic detection in the solar interior for
many years.

\begin{figure}[!t]
  \begin{center}
    \includegraphics{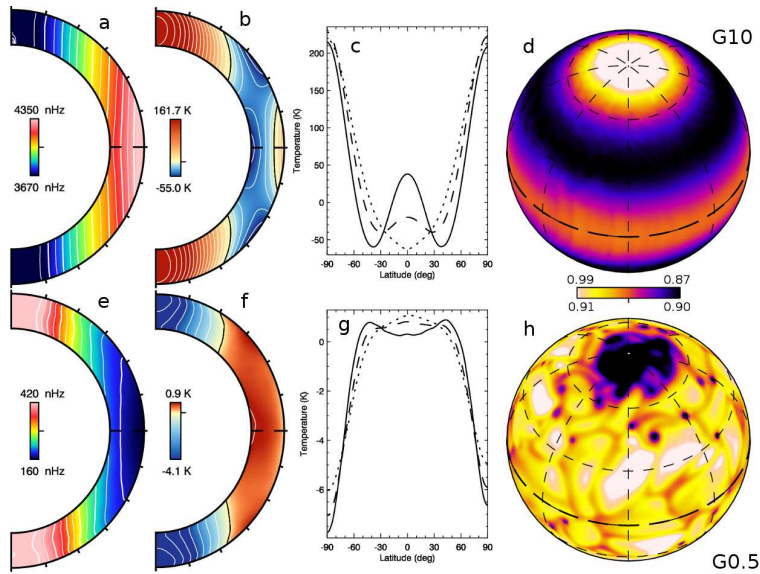}
  \end{center}
  \caption{Global-scale signatures of convection.  $(a)$ Angular
    velocity of differential rotation in rapidly-rotating case G10.
    The flow is solar-like, with a fast equator and slower flows at
    the poles.  The pole to equator contrast is about 10\% near the
    surface.  This zonal flow is substantially in thermal wind balance
    and the resulting temperature profile is shown in $(b)$ with cuts
    at constant radius shown in $(c)$ (solid near surface, dashed
    mid-convection zone, dotted bottom of convection zone).  
    $(d)$~Blackbody estimate of change in flux between polar and equatorial
    regions at the stellar photosphere, with hot north pole visible.  Scale is relative to
    stellar average flux and is clipped.
    In slowly rotating stars like case~G0.5, the sense of differential
    rotation may be anti solar $(e)$ with fast poles and slower
    equator.  The resulting thermal wind balance is cold at the poles
    and warm at the equator $(f,g)$ which may result in dark poles
    $(h)$ due to thermal effects rather than surface magnetism.  Here
    the surface flux difference is a few percent.
    \label{fig:signatures}}
\end{figure}

At the highest rotation rates however, surprising patterns of localized
convection emerge, and these self-organized structures may create
strong observational signatures.  Here flows near the equator may be confined to one
or two active ranges of longitude, with quiescent streaming flow in
between.  One such active nest of convection is shown in case~G10 in
Figure~\ref{fig:convection}$d$, with Figure~\ref{fig:convection}$e$
showing the entire near-surface layer in a global Mollweide view.
These active nests of convection are very long lived structures that
persist for thousands of days (many hundreds of convective turnover
times or rotation periods).  They move at their own angular velocity,
distinct from either the stellar rotation rate or the differential
rotation in which they are embedded and at times may cover a
substantial fraction of the stellar disk.  These structures have been found
in hydrodynamic simulations \citep{Brown_et_al_2008} and in some
situations they survive in the presence of magnetism.  In these cases
they can act to concentrate surface magnetism into localized
structures which may share many similarities with active longitudes of
stellar magnetism.

In all solar-like stars the coupling of rotation with convection
builds global-scale flows of differential rotation and meridional
circulation.  The profile of differential rotation is shown for
case~G10 in Figure~\ref{fig:signatures}$a$.  In the Sun and in more
rapidly rotating solar-type stars the angular velocity of differential
rotation is faster near the equator and slower near the polar
regions.  Thus the equator rotates prograde relative to the poles,
here with a relative contrast of about 10\% between the equator and
high-latitudes ($\pm60^\circ$). In contrast, the meridional
circulations are quite weak in these stars, decrease in amplitude with
faster rotation, and are generally multi-cellular in both radius and
latitude \citep{Brown_et_al_2008}.

The differential rotation is generally in a thermal-wind balance, and
this results in a latitudinal temperature structure that is hot near
the polar regions and cooler at mid-latitudes.  In the Sun the
magnitude of this contrast is probably only a few Kelvin and detecting this
signature has been very challenging \citep[e.g.,][]{Rast_et_al_2008}.
In more rapidly rotating stars the contrast may be much larger: here
in case~G10 a contrast of a few hundred Kelvin appears in the bulk of
the convection zone and likely prints through to the stellar surface 
(Fig.~\ref{fig:signatures}$b,c$). A very simplistic blackbody flux
estimate indicates that this may cause 10\% or larger differences
in the surface flux between the polar regions and the cooler mid-latitudes
(Fig.~\ref{fig:signatures}$d$).  
More details on thermal-wind balance in simulations of solar-type stars can be
found in \cite{Brun&Toomre_2002}, \cite{Miesch_2005}, \cite{Miesch_et_al_2006,Miesch_et_al_2008},
  \cite{Ballot_et_al_2007}, and \cite{Brown_et_al_2008}.

At slower rotation rates the dynamics may be substantially different
than in the rapidly rotating suns.  In particular, as the rotational
constraint weakens, the sense of differential rotation can flip and
become anti-solar, with fast poles and a slow equator.  This is shown
for case G0.5 in Figure~\ref{fig:signatures}$e$, where the poles
rotate nearly three times faster than the equator.  These slowly
spinning suns can remain in approximate thermal wind balance as well,
though again the sense flips, now with cool poles and warm equators
(Fig.~\ref{fig:signatures}$f,g$). 

A striking consequence of this is that the poles may be relatively
dark regions, due to fluid dynamic effects and irrespective of surface
magnetic structures there.  Here, under a simplistic blackbody flux
approximation, the pole is a few percent dimmer than the warmer
equator (Fig~\ref{fig:signatures}$h$).  Polar spots on slowly rotating
stars should be carefully examined to see whether any are due to
non-magnetic effects.  We note briefly that the designation 
``slowly spinning'' is dependent on spectral type, and brighter, luminous F-type stars will
be in the ``slow'' regime (high Rossby number) even if they rotate several times faster
than our Sun currently does \citep{Augustson_et_al_2011}.  Conversely, less-luminous K- and M-type
stars are likely ``rapid rotators'' (low Rossby numbers) even when spinning more slowly than the
Sun.  Dynamo solutions exist for the slowly spinning suns and these
retain an anti-solar differential rotation, though again of reduced
amplitude relative to the hydrodynamic simulations.

\section{Signatures of internal dynamics}

\begin{figure}
\begin{center}
  \includegraphics[width=\linewidth]{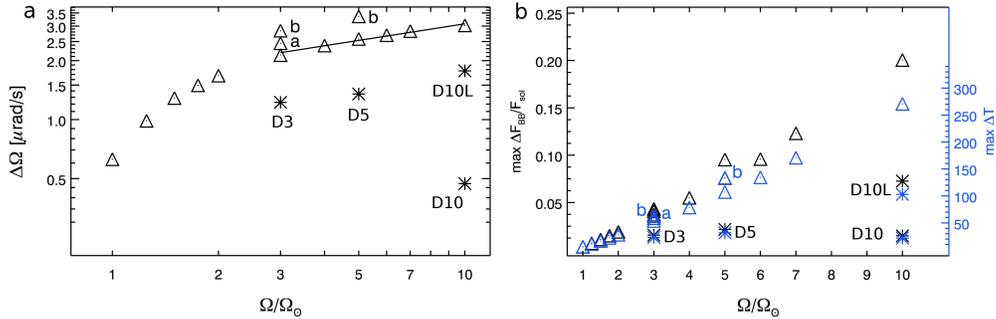}
\end{center}
\caption{Signatures of differential rotation and the thermal wind
  \citep[based on][]{Brown_et_al_2008}.  ($a$) Angular 
  velocity shear of differential rotation $\Delta \Omega$ in
  latitude near the stellar surface shown as a function of rotation
  rate $\Omega$ relative to the solar rotation rate
  $\Omega_\odot=2.6\mu\mathrm{rad/s}$.  Hydrodynamic cases are shown with diamonds while
  dynamos are labeled and shown with asterisks.  $\Delta \Omega$
  grows with more rapid rotation in hydrodynamic cases.  Cases labeled
  a and b sample more turbulent states.  ($b$)
  Thermal-wind signatures.  Shown in blue
  symbols (and right axis) are the largest latitudinal temperature
  contrasts achieved in these simulations.  Shown in black symbols
  (and left axis) is the maximum of the relative blackbody flux
  contrast between the bright poles and dim mid-latitudes.  
  \label{fig:scaling}
}
\end{figure}

The scaling of differential rotation and thermal-wind for the rapidly
rotating simulations are shown in Figure~\ref{fig:scaling}.  Generally, we find
that the angular velocity shear between equator and high latitudes
$\Delta \Omega$ grows with faster rotation, though not as quickly
as the rotation rate itself (a power-law of $\Delta \Omega \propto
\Omega^{0.3}$ is overplotted on cases rotating faster than $3\:\Omega_\odot$).  
Owing to this, the relative angular velocity contrast
\begin{equation}
  \Delta \Omega/\Omega = \Big({\Omega(\mathrm{equator})-\Omega(\mathrm{pole})}\Big)/{\Omega(\mathrm{equator})}
\end{equation}
decreases as the rotation rate increases,
here scaling as $\Omega^{-0.6}$ \citep{Brown_et_al_2008}.  
When magnetic fields are included and dynamo simulations are
conducted, the global-scale magnetic fields weaken the large scale
differential rotation (Fig.~\ref{fig:scaling}, asterisks).  
At present we are still sorting out how magnetism and differential
rotation couple, and simulations right now are suggesting that the
answer may change in different regimes of parameter space
\citep[e.g., cases D10 and D10L in Fig.~\ref{fig:scaling}$a$, and
  see][]{Brown_2011}. 

The thermal-wind balance leads to progressively larger latitudinal
contrasts of temperature in more rapidly rotating simulations
(Fig.~\ref{fig:scaling}$b$).  In the most rapidly rotating cases, the
hot poles and the cool mid-latitudes can be nearly 300K different in
temperature.  If this latitudinal temperature gradient prints through
the vigorous surface convection, then simple blackbody flux arguments
would suggest that the poles could be up to 20\% brighter than the
cool mid-latitude bands (Fig.~\ref{fig:scaling}$b$, black symbols).
The magnetic fields in the dynamo simulations reduce this signature as
they reduce the differential rotation, and in those simulations the
signature is about 5--10\%.  These blackbody fluxes are far too simple and
neglect nearly all aspects of the stellar atmosphere, including
viewing angle, but provide an order of magnitude initial estimate.

\section{Wreath-building dynamos}
The magnetic fields produced in these dynamo simulations may also have
observational consequences. Generally, these rapidly rotating suns
generate strong magnetic fields in the bulk of their convection zones.  
This is surprising as many solar dynamo theories hold that such
organized dynamo action can only occur in the tachocline of shear and
penetration located between the base of the convection zone and the
stable radiative zone below.  These simulations do not include that
interface layer yet they still build organized fields.

\begin{figure}[!t]
  \begin{center}
    \includegraphics[width=\linewidth]{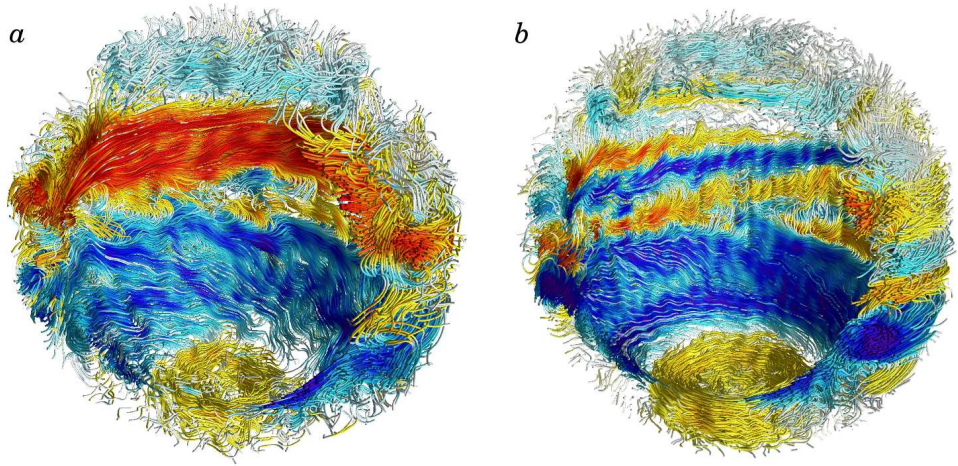}
  \end{center}
  \caption{Magnetic wreaths in stellar convection zones. 
    $(a)$ Magnetic wreaths in cyclic case D5.  Two wreaths of opposite
    polarity (red, positive; blue, negative) form above and below the equator.
    In this simulation the wreaths undergo reversals of polarity on
    roughly a 1500-day timescale.  Relic wreaths from the previous
    cycle are visible in  the polar caps.  
    $(b)$ Wreaths in more-rapidly rotating case D10L.  Here a negative
    polarity axisymmetric wreath dominates the southern hemisphere
    while the northern hemisphere is filled with non-axisymmetric
    fields.  During the next cycle the roles reverse and the wreath
    instead appears in the northern hemisphere, with tangled fields in
    the southern.  Rings of relic fields are strongly evident at the
    southern pole. In both images the colortable saturates at $\pm
    25$kG, while the fields may reach peak amplitudes of more than $\pm 50$kG.
    \label{fig:magnetic wreaths D5 D10L}    
  }
\end{figure}

Magnetic fields from two of the rapidly rotating dynamo cases are
shown in Figure~\ref{fig:magnetic wreaths D5 D10L}. These magnetic
fields are organized in global-scale wreath-like structures, with
complex topologies.  Magnetic cycles are achieved in many of these
dynamo simulations, including both cases D5 and D10L shown here.  
In some cases (e.g, D5 in Fig.~\ref{fig:magnetic wreaths D5 D10L}$a$) 
the wreaths are highly axisymmetric and have opposite polarities in
each hemisphere.  During the magnetic cycle, wreaths of opposite
polarity form in each hemisphere and at a later time the polarities
will have reversed.  Roughly 1500 days later the cycle repeats and
returns the magnetic fields to a state like is shown here.  
In some cases (e.g, D10L in Fig.~\ref{fig:magnetic wreaths D5 D10L}$b$)
the wreaths may be much more concentrated in one hemisphere (here the
southern).  This simulation shows cycles of activity and in each
successive cycle the wreaths alternate between the northern and
southern hemisphere, though they rarely fill both at the same time.
Magnetic fields are present in the other hemisphere (here the
northern) but are less axisymmetric.  In both case D5 and D10L there
are significant rings of opposite polarity field located at the polar
caps.  These are relic fields from the preceding activity cycle
\citep{Brown_et_al_2011}.

\section{Constraining simulations with interferometric observations}

Modern simulations of convection in solar-type stars are able to
self-consistently generate global-scale flows of differential rotation
and meridional circulation.  In the simulations, the convection and
differential rotation drive strong dynamo action in the bulk of the
stellar convection zone, generating coherent global-scale wreaths of
magnetism. 
The zonal flow of differential rotation is likely the most easily
observable signature of dynamics occuring deep within the stellar interior.
Its characterization is thus of crucial importance.  
Despite this, current observations are in significant
disagreement \citep[e.g.,][]{Donahue_et_al_1996, Barnes_et_al_2005}, 
and there are hints of very interesting relationships between $\Delta
\Omega$ and the X-ray luminosity of stars \citep{Saar_2009}.  

Simulations predict that the angular velocity contrast in latitude
should grow larger in solar-type stars as they rotate more rapidly. At the highest rotation
rates, the relative contrast $\Delta \Omega/\Omega \sim 10\%$, which 
at ten times the solar rotation rate (case~G10, with rotation period
of about 3~days) would lead to an equatorial velocity of about 20~km/s
and a high-latitude ($60^\circ$ latitude) velocity of about 14~km/s 
(versus 17~km/s for solid body rotation).  These are clearly
challenging observations but may lie within reach for current
interferometers (e.g., VEGA at CHARA).  To inspire such searches, 
a few of the nearby, rapidly-rotating and bright solar-type stars
which should be accessible to northern-hemisphere interferometers are
listed in Table~\ref{table:targets}. Rotation rates here are lower
limits, based on Vsini from SIMBAD and rough estimates of stellar
radii from \cite{Fracassini_et_al_2001}.

The differential rotation may be in thermal wind balance and this may
lead to latitudinal gradients of temperature.  These may in term lead
to gradients of the stellar flux equal to a few percent in
brightness.  Solar-like differential rotation is probably accompanied
by relatively warm poles and cool mid-latitudes, while anti-solar
differential rotation would have cool poles and warm equators.  These
thermal signatures are non-magnetic in nature.

\begin{table}[!t]
\caption{A few possible observational targets\label{table:targets}}
\smallskip
\begin{center}
{\small
\begin{tabular}{cccccccc}
\tableline
\noalign{\smallskip}
Name & HD & Spectral type & parallax & Vmag & Vrot & $R/R_\odot$ & $\Omega/\Omega_\odot$ \\
\noalign{\smallskip}
\tableline
\noalign{\smallskip}
    tet Boo    & 126660\phantom{A} &  F7V\phantom{I}  &    68   & 4.1 &   32 & 1.5 & 10\\
    chi Dra    & 170153\phantom{A} &  F7V\phantom{I}  &   124   & 3.6 &    8 & 1.1 & 4 \\
    86 Her     & 161797\phantom{A} &  G5IV &   119    &   3.4   &  8 & 1.5 & 3 \\
    GJ 702 A   & 165341A  &  K0V\phantom{I}  &   196  &   4.2 &   16 & 0.8 & 10 \\
\noalign{\smallskip}
\tableline
\end{tabular}
}
\end{center}
\end{table}

Lastly, the global-scale longitudinal and radial magnetic fields associated
with these wreaths may appear at or near the stellar surface.  Observations
with Zeeman-Doppler Imaging (ZDI) techniques appear to show evidence
for large-scale longitudinal magnetic fields at or near the stellar surfaces
\citep[e.g.,][]{Petit_et_al_2008, Marsden_et_al_2011a}.
Interferometric spectroscopic observations may be able to spatially resolve these
structures, providing critical tests for both the simulations and the
growing field of ZDI observations, which can in principle be applied
to more distant objects.  Characterizing the large-scale poloidal
field may also provide indirect indications of large wreaths lurking beneath
the stellar photospheres.

\vspace{0.5truein}

\acknowledgements  This research on wreath-building dynamos in
solar-type stars has been done in collaboration with Kyle
C.\ Augustson, Matthew K.\ Browning, Allan Sacha Brun, 
Mark S.\ Miesch, Nicholas J.\ Nelson and Juri Toomre, and I owe them many thanks.  
Funding for this research is provided in part through NSF Astronomy
and Astrophysics Postdoctoral Fellowship AST 09-02004.  CMSO is
supported by NSF grant PHY 08-21899.  The simulations were carried out
with NSF PACI support of NICS, PSC and TACC.   
Field line tracings shown in Figure~\ref{fig:magnetic wreaths D5 D10L} 
were produced using VAPOR \citep{Clyne_et_al_2007}.
This research has made use of the SIMBAD database,
operated at CDS, Strasbourg, France 

\newpage
\bibliographystyle{asp2010}
\bibliography{bibliography}

\begin{thebibliography}{}
\expandafter\ifx\csname natexlab\endcsname\relax\def\natexlab#1{#1}\fi
\expandafter\ifx\csname url\endcsname\relax
  \def\url#1{\texttt{#1}}\fi
\expandafter\ifx\csname urlprefix\endcsname\relax\def\urlprefix{URL }\fi
\providecommand{\eprint}[2][]{\url{#2}}

\bibitem[{{Augustson} et~al.(2011){Augustson}, {Brown}, {Brun}, \&
  {Toomre}}]{Augustson_et_al_2011}
{Augustson}, K., {Brown}, B.~P., {Brun}, A.~S., \& {Toomre}, J. 2011, \apj, in
  preparation

\bibitem[{{Ballot} et~al.(2007){Ballot}, {Brun}, \&
  {Turck-Chi{\`e}ze}}]{Ballot_et_al_2007}
{Ballot}, J., {Brun}, A.~S., \& {Turck-Chi{\`e}ze}, S. 2007, \apj, 669, 1190.
  \eprint{arXiv:0707.3943}

\bibitem[{{Barnes} et~al.(2005){Barnes}, {Cameron}, {Donati}, {James},
  {Marsden}, \& {Petit}}]{Barnes_et_al_2005}
{Barnes}, J.~R., {Cameron}, A.~C., {Donati}, J.-F., {James}, D.~J., {Marsden},
  S.~C., \& {Petit}, P. 2005, \mnras, 357, L1. \eprint{arXiv:astro-ph/0410575}

\bibitem[{{Brown}(2011)}]{Brown_2011}
{Brown}, B. 2011, Journal of Physics Conference Series, 271, 012064.
  \eprint{1011.2992}

\bibitem[{{Brown} et~al.(2008){Brown}, {Browning}, {Brun}, {Miesch}, \&
  {Toomre}}]{Brown_et_al_2008}
{Brown}, B.~P., {Browning}, M.~K., {Brun}, A.~S., {Miesch}, M.~S., \& {Toomre},
  J. 2008, \apj, 689, 1354. \eprint{0808.1716}

\bibitem[{{Brown} et~al.(2010){Brown}, {Browning}, {Brun}, {Miesch}, \&
  {Toomre}}]{Brown_et_al_2010a}
--- 2010, \apj, 711, 424. \eprint{1011.2831}

\bibitem[{{Brown} et~al.(2011){Brown}, {Miesch}, {Browning}, {Brun}, \&
  {Toomre}}]{Brown_et_al_2011}
{Brown}, B.~P., {Miesch}, M.~S., {Browning}, M.~K., {Brun}, A.~S., \& {Toomre},
  J. 2011, \apj, 731, 69:1. \eprint{1102.1993}

\bibitem[{{Browning} et~al.(2006){Browning}, {Miesch}, {Brun}, \&
  {Toomre}}]{Browning_et_al_2006}
{Browning}, M.~K., {Miesch}, M.~S., {Brun}, A.~S., \& {Toomre}, J. 2006, \apjl,
  648, L157. \eprint{arXiv:astro-ph/0609153}

\bibitem[{{Brun} et~al.(2004){Brun}, {Miesch}, \& {Toomre}}]{Brun_et_al_2004}
{Brun}, A.~S., {Miesch}, M.~S., \& {Toomre}, J. 2004, \apj, 614, 1073

\bibitem[{{Brun} \& {Toomre}(2002)}]{Brun&Toomre_2002}
{Brun}, A.~S., \& {Toomre}, J. 2002, \apj, 570, 865.
  \eprint{arXiv:astro-ph/0206196}

\bibitem[{{Clune} et~al.(1999){Clune}, {Elliott}, {Glatzmaier}, {Miesch}, \&
  {Toomre}}]{Clune_et_al_1999}
{Clune}, T.~L., {Elliott}, J.~R., {Glatzmaier}, G.~A., {Miesch}, M.~S., \&
  {Toomre}, J. 1999, Parallel Computing, 25, 361

\bibitem[{{Clyne} et~al.(2007){Clyne}, {Mininni}, {Norton}, \&
  {Rast}}]{Clyne_et_al_2007}
{Clyne}, J., {Mininni}, P., {Norton}, A., \& {Rast}, M. 2007, New Journal of
  Physics, 9, 301

\bibitem[{{Donahue} et~al.(1996){Donahue}, {Saar}, \&
  {Baliunas}}]{Donahue_et_al_1996}
{Donahue}, R.~A., {Saar}, S.~H., \& {Baliunas}, S.~L. 1996, \apj, 466, 384

\bibitem[{{Ghizaru} et~al.(2010){Ghizaru}, {Charbonneau}, \&
  {Smolarkiewicz}}]{Ghizaru_et_al_2010}
{Ghizaru}, M., {Charbonneau}, P., \& {Smolarkiewicz}, P.~K. 2010, \apjl, 715,
  L133

\bibitem[{{Marsden} et~al.(2011){Marsden}, {Jardine}, {Ram{\'{\i}}rez
  V{\'e}lez}, {Alecian}, {Brown}, {Carter}, {Donati}, {Dunstone}, {Hart},
  {Semel}, \& {Waite}}]{Marsden_et_al_2011a}
{Marsden}, S.~C., {Jardine}, M.~M., {Ram{\'{\i}}rez V{\'e}lez}, J.~C.,
  {Alecian}, E., {Brown}, C.~J., {Carter}, B.~D., {Donati}, J.-F., {Dunstone},
  N., {Hart}, R., {Semel}, M., \& {Waite}, I.~A. 2011, \mnras, 413, 1922.
  \eprint{1101.5859}

\bibitem[{{Miesch}(2005)}]{Miesch_2005}
{Miesch}, M.~S. 2005, Living Reviews in Solar Physics, 2, 1:1

\bibitem[{{Miesch} et~al.(2008){Miesch}, {Brun}, {DeRosa}, \&
  {Toomre}}]{Miesch_et_al_2008}
{Miesch}, M.~S., {Brun}, A.~S., {DeRosa}, M.~L., \& {Toomre}, J. 2008, \apj,
  673, 557. \eprint{arXiv:0707.1460}

\bibitem[{{Miesch} et~al.(2006){Miesch}, {Brun}, \&
  {Toomre}}]{Miesch_et_al_2006}
{Miesch}, M.~S., {Brun}, A.~S., \& {Toomre}, J. 2006, \apj, 641, 618

\bibitem[{{Miesch} \& {Toomre}(2009)}]{Miesch&Toomre_2009}
{Miesch}, M.~S., \& {Toomre}, J. 2009, Annual Review of Fluid Mechanics, 41,
  317

\bibitem[{{Pasinetti Fracassini} et~al.(2001){Pasinetti Fracassini}, {Pastori},
  {Covino}, \& {Pozzi}}]{Fracassini_et_al_2001}
{Pasinetti Fracassini}, L.~E., {Pastori}, L., {Covino}, S., \& {Pozzi}, A.
  2001, \aap, 367, 521. \eprint{arXiv:astro-ph/0012289}

\bibitem[{{Petit} et~al.(2008){Petit}, {Dintrans}, {Solanki}, {Donati},
  {Auri{\`e}re}, {Ligni{\`e}res}, {Morin}, {Paletou}, {Ramirez Velez},
  {Catala}, \& {Fares}}]{Petit_et_al_2008}
{Petit}, P., {Dintrans}, B., {Solanki}, S.~K., {Donati}, J.-F., {Auri{\`e}re},
  M., {Ligni{\`e}res}, F., {Morin}, J., {Paletou}, F., {Ramirez Velez}, J.,
  {Catala}, C., \& {Fares}, R. 2008, \mnras, 388, 80. \eprint{0804.1290}

\bibitem[{{Pizzolato} et~al.(2003){Pizzolato}, {Maggio}, {Micela}, {Sciortino},
  \& {Ventura}}]{Pizzolato_et_al_2003}
{Pizzolato}, N., {Maggio}, A., {Micela}, G., {Sciortino}, S., \& {Ventura}, P.
  2003, \aap, 397, 147

\bibitem[{{Racine} et~al.(2011){Racine}, {Charbonneau}, {Ghizaru}, {Bouchat},
  \& {Smolarkiewicz}}]{Racine_et_al_2011}
{Racine}, {\'E}., {Charbonneau}, P., {Ghizaru}, M., {Bouchat}, A., \&
  {Smolarkiewicz}, P.~K. 2011, \apj, 735, 46

\bibitem[{{Rast} et~al.(2008){Rast}, {Ortiz}, \& {Meisner}}]{Rast_et_al_2008}
{Rast}, M.~P., {Ortiz}, A., \& {Meisner}, R.~W. 2008, \apj, 673, 1209.
  \eprint{0710.3121}

\bibitem[{{Saar}(2009)}]{Saar_2009}
{Saar}, S.~H. 2009, in Solar-Stellar Dynamos as Revealed by Helio- and
  Asteroseismology: GONG 2008/SOHO 21, edited by {M.~Dikpati, T.~Arentoft,
  I.~Gonz{\'a}lez Hern{\'a}ndez, C.~Lindsey, \& F.~Hill}, vol. 416 of
  Astronomical Society of the Pacific Conference Series, 375

\end{thebibliography}

\end{document}